\documentclass[draft]{agujournal2019}
\usepackage{url} 
\usepackage{lineno}
\usepackage[inline]{trackchanges} 
\usepackage{soul}
\draftfalse

\journalname{Geophysical Research Letters}

\begin{document}

\title{VLF Remote Sensing across and along the Totality Path during the April 8, 2024 Total Solar Eclipse using a VLF Receiver Network}


\authors{Oleksiy Agapitov\affil{1,5}, Mark Go{\l}kowski\affil{2}, Lucas Colomban\affil{1}, Ryan Eskola\affil{2}, Favour Ogbinaka\affil{2}, Iryna Agapitova\affil{1}, Kyung-Eun Choi\affil{1}, Varvara Bashkirova\affil{3}, Aaron Brenemean\affil{4}}
\affiliation{1}{Space Sciences Laboratory, University of California Berkeley, Berkeley, CA, USA}
\affiliation{2}{Department of Electrical Engineering, University of Colorado Denver, Denver, CO, USA}
\affiliation{3}{University of Michigan Ann Arbor, Ann Arbor, MI, USA}
\affiliation{4}{NASA Goddard Space Flight Center, MD, USA}
\affiliation{5}{Astronomy and Space Physics Department, National Taras Shevchenko University of Kyiv, Kyiv, Ukraine}

\correspondingauthor{Oleksiy Agaptiov}{agaptiov@ssl.berkeley.edu}

\begin{keypoints}
\item Novel radio observations of the total solar eclipse on April 8, 2024, by a network of VLF Receivers deployed along and across the Totality Path
\item Observations show unique features related to the percentage of obscurity over the receiver, over the transmitter, and along the radio path. 
\item Unique quasi-periodic `W' signature of moon shadow traversing the radio path is observed for the first time. 
\end{keypoints}

\begin{abstract}

During the total solar eclipse in the United States on April 8, 2024, we observed the amplitude and phase of VLF signals from five U.S. Navy VLF transmitters using a novel geometry of VLF receivers deployed along and across the totality path. Nine receiver sites (four of them inside the totality path) were deployed to collect the data in different transmitter-receiver configurations relative to the totality path, which intersected with the radio propagation paths from the Cutler, MA (NAA, 24 kHz) and the  LaMoure, ND (NML, 25.2 kHz) Navy transmitters. The transmitters themselves experienced  98.7\%  (NAA) and  68\% (NML) solar obscuration at 75 km altitude. The novelty of the observations is the near-total obscuration of one of the transmitters and observations of several radio propagation paths closely aligned to the path of totality. This configuration enabled observations of the effects of the Moon’s shadow progression from Texas to Maine for over three hours on a total of 28 radio paths (with coverage from 55\% to 100\%) with transmitter-to-receiver distances ranging from 780 km to 7700 km.  Both positive and negative (5- 10 dB) amplitude changes were observed by receivers throughout the eclipse period. The observed phase changes were mostly negative. The unique observations of VLF propagation along the totality path produced a temporally dynamic quasiperiodic `W' shaped response in amplitude that can be used to determine the gradients and spatial scales of the eclipse effect on the lower ionosphere.  For observations within 780 km of NAA, the eclipse produced a 13 dB surge in amplitude. 
\end{abstract}

\section*{Plain Language Summary}
Solar eclipses provide a unique way to study day-night changes in the upper atmosphere, known as the ionosphere.  Very low frequency (VLF) waves from powerful navy transmitters are sensitive to the changes in the upper atmosphere, and their observations are an established way of remotely sensing the lower ionosphere.  During the total solar eclipse in the USA on April 8, 2024, VLF signal measurements were taken from five U.S. Navy transmitters. A network of receivers was set up along the eclipse path, with seven sites (four within the totality zone). This configuration provided comprehensive data, with a particular focus on the Cutler, Maine (NAA) transmitter, which experienced 99.7\% sun blockage. Data from 28 radio paths, spanning coverage levels from 55\% to 100\%, were gathered, capturing signal variations over distances ranging from 780 km to 7700 km. One receiver was in the location of maximum totality duration, and three were on the edge. Both increases and decreases in signal amplitude were observed. The unique setup with the Cutler transmitter and receivers along the totality path enabled a detailed analysis of the D-region's response, providing information on the temporal and spatial characteristics of these atmospheric changes.

%
%

\section{Introduction}

Solar eclipses provide rare opportunities to probe the ionospheric response to rapid changes in solar radiation. The sudden reduction in the ionizing flux of solar irradiation during an eclipse creates a transient ionospheric twilight, with the complex interplay of ionization/recombination and photochemical processes, neutral dynamics, and wave propagation. The study of the ionospheric response to solar eclipses has a long history \cite{rishbeth1968solar,Tsai1999,Afraimovich2002,Baran2003,Jakowski2008, Xiong2023, Resende2021}, and various remote sensing techniques, such as GPS, ionosondes, and satellite radio beacons, have been employed to analyze Total Electron Content (TEC) and other parameters \cite{Momani2010, Hoque2016,Zhang2017,Dang2020,Chen2021} including the eclipse discussed here \cite{gautam_ionospheric_2024}.

Although all regions of the ionosphere are influenced by solar radiation, the lowest portion, the D-region (altitudes of 60$-$90 km), undergoes the most dramatic changes in response to its presence or absence. This effect is evident in the commonly held, though greatly oversimplified,  belief that the D-region `disappears' at night \cite{golkowski2018ionospheric}. At all times, the D-region plays a crucial role in governing the propagation of ELF/VLF waves, and the absorption of HF waves, and is highly sensitive to energetic particle precipitation \cite{helliwell1973whistler, golkowski2014observation} from the Earth's magnetosphere, as well as solar flares \cite{nieckarz2024monitoring, ostrowski2024effects} and gamma-rays  \cite{inan2007massive}. Consequently, the lower ionosphere is of interest to a diverse range of researchers. However, it presents significant challenges since direct measurements of electron density are difficult to obtain, and radio remote sensing often fails to yield a unique solution \cite{golkowski2021quantification, gross_vlf_2020}.

The most prevalent approach to remote sensing of the lower ionosphere is using narrowband VLF  waves from military communication transmitters, and such techniques have been used to observe solar eclipse effects since at least 1949 \cite{bracewell_theory_1952, crary1965effect,albee_vlf_1965,decaux1964some, kaufmann_effect_1968,schaal_vlf_1970}. Since VLF propagation in the Earth-ionosphere waveguide is multimodal, the observed perturbations in amplitude and phase are dependent on the specifics of the radio propagation path, making multi-point measurements critical for conclusive observations. \citeA{clilverd_total_2001}  reported one of the first multi-receiver observations of VLF transmitter signals in the range 16-24 kHz during an eclipse: Five receiver sites were set to receive signals from four transmitters located in Europe during the eclipse of August 11, 1999. The results confirmed the general tendency for an increase in signal amplitude on paths of lengths less than 2000 km. Negative amplitude changes were observed on paths L $>$ 10,000 km. Negative phase changes were observed on most paths, independent of path length. Although there was significant variation from path to path, the typical changes observed were $\sim$3 dB and $\sim$50$^o$ \cite{clilverd_total_2001}. The 2017 total solar eclipse in North America saw several targeted VLF observations that confirmed previous work and also suggested new areas of focus \cite{cohen_lower_2018, xu_vlf_2019, chakrabarti2018modeling, rozhnoi_effect_2020}. 

A shortcoming of previous work has been that the size and gradients of the eclipse-induced perturbation have not been spatially resolved. Many of the past works have been able to reproduce their observations to various degrees by assuming gradual decreases in electron density in the entire sun-lit ionosphere \cite{clilverd_total_2001} or have assumed that the vertical electron density gradients remain constant \cite{xu_vlf_2019}. At the same time, there have been suggestions in previous work that the gradients near the totality shadow can be significant. 
In particular, \citeA{cohen_lower_2018} showed evidence of direct reflective scattering off the narrow 100-km-wide totality spot when the transmitter and receiver were relatively close to the totality spot. \citeA{xu_vlf_2019} also estimated the effective size of the eclipse-induced perturbation. It is important to note that the totality zone is the optical shadow, and X-ray emissions from the solar corona can continue to drive D-region ionization even in totality conditions. In this context, \cite{xu_vlf_2019} presented maps of  far-/mid-ultraviolet (FUV/MUV) and extreme ultraviolet (EUV) wavelength obscuration. 

The present work is one of the first to present multiple observations of the eclipse perturbation along the same radio propagation path. This is enabled by a transmitter-receiver geometry that is along the totality zone, allowing for continuous observation of the shadow progression along the path.  Furthermore, we present the results in the context of the obscuration of the receiver location, the transmitter location, and the average obscuration over the radio propagation path.

On April 8, 2024, a total solar eclipse occurred over North America with a maximum duration of totality of 4 minutes and 27 seconds. The eclipse path in the United States (shown in Figure \ref{Fig1}) provided a unique opportunity to deploy a network of VLF receivers and collect data along the totality path. The 0.99 coverage of the Sun at Cutler, Maine - the location of the NAA Navy transmitter - was particularly beneficial for the experiment. We present the results of observations of the signals from the five US Navy VLF transmitters collected by eight VLF receiver stations (3 from the University of Colorado Denver and 5 from the University of California Berkeley) deployed in Denver, CO; San Antonio, TX; Austin, TX; Antlers, OK; Cleveland, OH; Ithaca, NY; College Park, MD; and Ann Arbor, MI.

\section{Observation Geometry and Hardware Deployment}

Two different types of VLF receivers were used to make VLF observations. The University of Colorado Denver (CU Denver) deployed receivers of the type described by \citeA{cohen2009sensitive} in Denver, CO; Antlers, OK; and Ithaca, NY. The CU Denver receivers have automatic demodulation to extract the carrier phase \cite{gross2018polarization}. The University of California, Berkeley (UCB) hardware was originally developed to support the NASA-guided VIPER rocket \cite{bonnell_first_2021} (the characteristics are presented in the Supplemental Material), and these systems were deployed in San Antonio, TX, Austin, TX, Antlers, OK, Cleveland, OH, College Park, MD, and Ann Arbor, MI. The specific locations of all deployed VLF receivers are listed in Table 1. Both receiver types record two orthogonal components of the horizontal magnetic fields of VLF waves.  {The CU Denver and UCB receivers were deployed in Antlers, OK for cross-correlation purposes and showed identical changes in observed amplitude (Figure\ref{Fig2}f). }

\begin{figure}[ht!]
 \includegraphics[width=\textwidth]{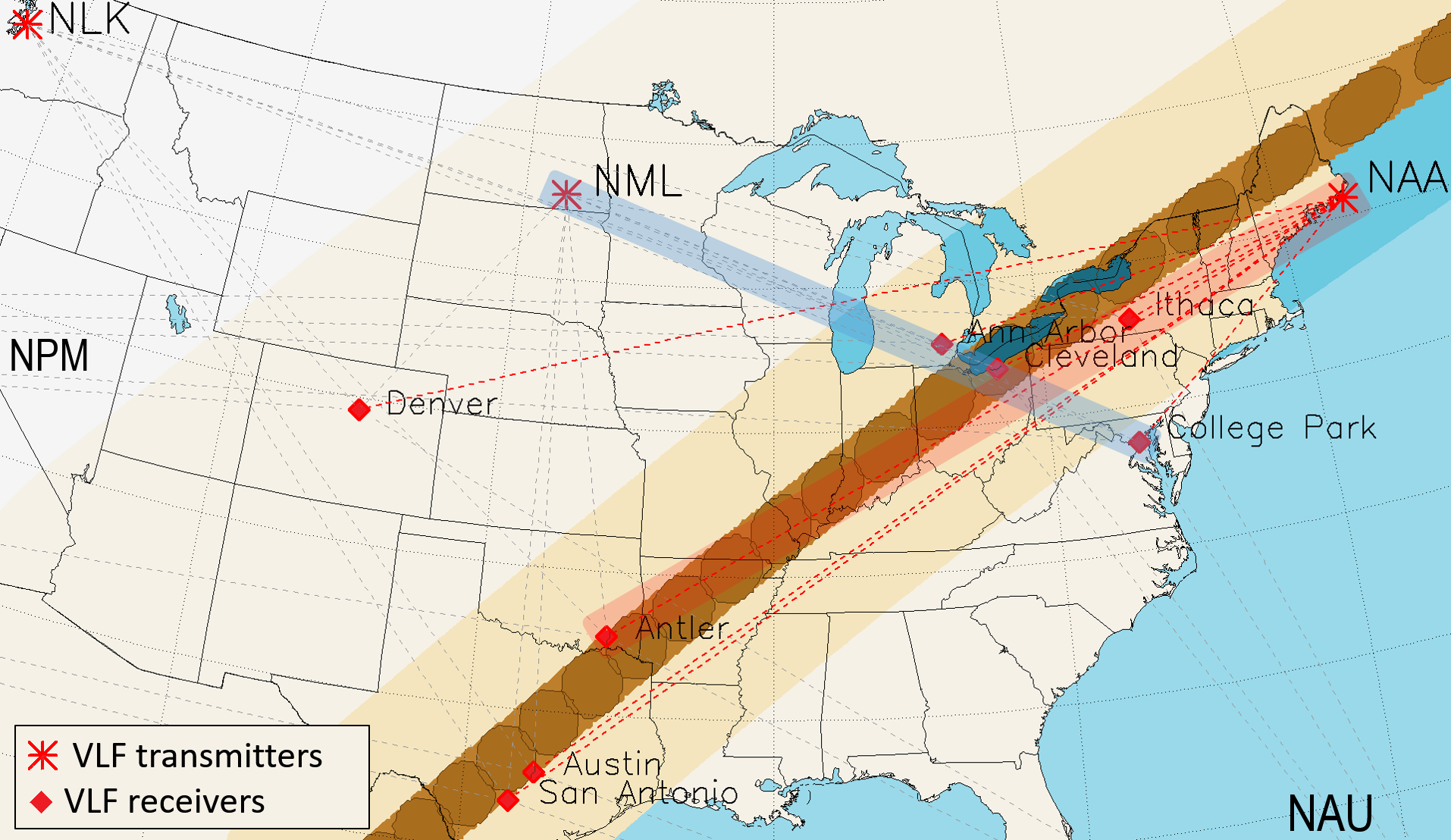}
 \caption{The eclipse totality path on April 8, 2024, calculated for 75 km altitude. The areas with 90\% and 50\% obscuration are highlighted in yellow and light yellow, respectively. The locations of the continental VLF transmitters NAA, NML, and NLK are marked with stars. The VLF receiver locations are indicated with red diamonds. The radio path from NAA to Antlers, OK, through Ithaca, NY, and Cleveland, OH, is highlighted in red. The path from NML to College Park, MD, through Ann Arbor, MI, and Cleveland, OH, is highlighted in blue.}
 \label{Fig1}
\end{figure}

\begin{figure}[ht!]
\center{\includegraphics[width=.99\textwidth]{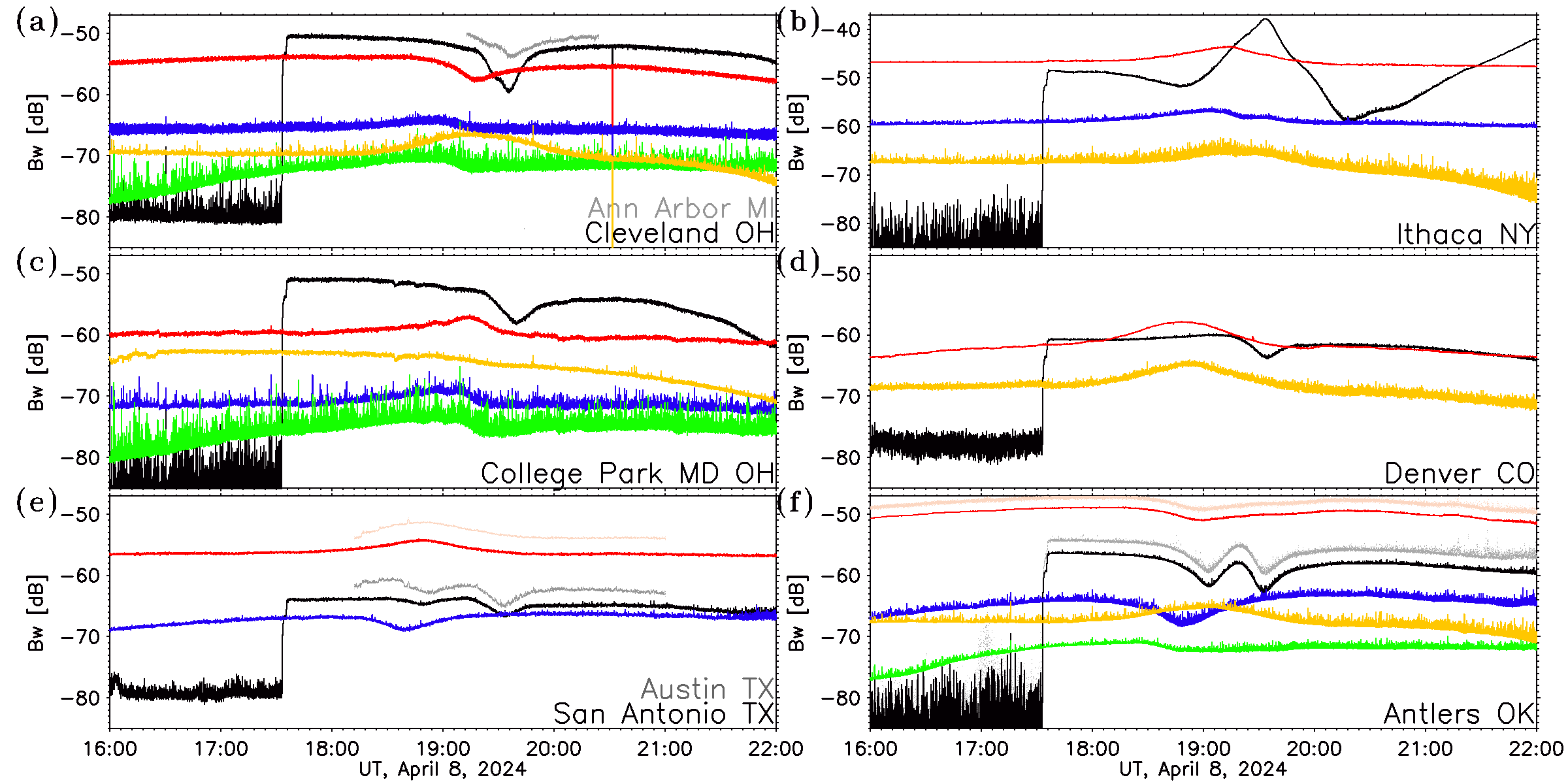}
 \caption{Dynamics of VLF signal at the frequencies of the five VLF transmitters (NAA - black, NML - red, NLK - blue, NPM - green, NAU - yellow) from the data collected by VLF receivers during 6 hours of April 8, 2024 (a) in Cleveland, OH and Ann Arbor, MI (NAA recording in Ann Arbor is indicated with the gray curve), (b) in Ithaca, NY, (c) in College Park MD, (d) in Denver, CO, (e) in San Antonio and Austin, TX (indicated with the light red and gray curves for NML and NAA signals respectively), and (f) in Antlers, OK (the measurements made by the UCB receiver are indicated with the light red and gray curves for NML and NAA signals respectively). 
 }}
 \label{Fig2}
\end{figure}

Figure \ref{Fig1} shows the location of the receivers relative to the major VLF transmitters and to the totality path at 75 km altitude. 
Following the recommendations of \citeA{golkowski2021quantification}, the receivers were set up on two primary radio propagation paths, perpendicular and parallel to the totality path.  Specifically, two lines (from NAA to Ithaca, NY, Cleveland, OH, Antlers, OK - highlighted in red in Figure \ref{Fig1}; and from NML to Ann Arbor MI, Cleveland, OH, College Park, MD - highlighted in blue in Figure \ref{Fig1}) were formed. Five VLF receivers were deployed on the edge of totality (San Antonio, TX; Austin, TX; Antlers, OK; Ann Arbor, MI; and Ithaca, NY) while one was placed in the region of maximum totality duration (3 minutes and 45 seconds) in Cleveland, OH. The VLF receivers monitored the amplitude and phase of the signal from the NAA Navy VLF transmitter ($44.65^\circ\mathrm{N}, 67.28^\circ\mathrm{W}$, 24.0 kHz, $\sim$ 1 MW);  the NLK transmitter ($48.20^\circ\mathrm{N}, 121.92^\circ\mathrm{W}$, 24.8 kHz, $\sim$ 250 kW);  the NAU transmitter ($18.38^\circ\mathrm{N}, 67.18^\circ\mathrm{W}$, 40.75 kHz, $\sim$100 kW); the NML transmitter ($46.36^\circ\mathrm{N}, 98.34^\circ\mathrm{W}$, 25.2 kHz, $\sim$500 kW); and the NPM transmitter ($21.42^\circ\mathrm{N}, 158.15^\circ\mathrm{W}$, 21.4 kHz, $\sim$400 kW).  We present the results of processing the 28 traces listed in Table 1, with the main focus on the 8 radio propagation paths from NAA because of their location on the edge of the totality and the 98\% obscuration of the transmitter location which occurred at 19:33 UT. 

\begin{table}[h]
\centering
\begin{tabular}{|l|c|c|c|c|c|c|}
\hline
 {Location and Obscuration}  &  {NAA} &  {NLK} &  {NPM} &  {NML} &  {NAU} \\
                                   & 44.6,-67.3 & 47.5,-122.3 & 21.4,-158.0 & 46.37,-98.34& 18.4,-67.2 \\ \hline
San Antonio$^{**}$(29.42,-98.49) 1.0& 3209(--/--)         & 2774(--)         & -         & 1890(+)         & -         \\ \hline
Austin TX$^{**}$ (30.29, -97.74) 1.0 & 3092(--/--)         & -         & -         & 1788(+)         & -        \\ \hline
Antlers OK$^*$ (34.21, -95.61) 1.0 & 2671(--/--)         & 2666(--)         & 6205(+/--)         & 1370(--)         & 3318(+)         \\ \hline
Denver$^*$ (39.69, -105.09) 0.71 & 3547(+/--)         & -         & -         & 922(+)         & 4333(+)         \\ \hline
College Park (39.01, -76.84) 0.89  & 998(nc/--)         & 3743(+)         & 7791(+/--)         & 1929(+)         & 2476(nc)         \\ \hline
Ann Arbor$^{**}$ (42.31, -83.69) 0.98  & 1343(--/--)         & -         & -         & -         & -      \\ \hline
Cleveland(41.45, -81.69) 1.0  & 1215(--/--)         & 3250(+)         & 7337(+/--)         & 1434(+)         & 2917(+)         \\ \hline
Ithaca$^*$ (42.42, -76.53) 0.98 & 780(--/+)         & 3590(+)         & -         & 1780(+)         & 2815(+)         \\ \hline
\end{tabular}


\caption{Distances in kilometers between the transmitters and the VLF receivers. The receivers' coordinates and the maximal obscuration of the receiver location are listed. c$^{*}$ - the signal carrier phase  is available, $^{**}$ - only one magnetic component is available. '+' or '-' indicates a positive or negative change in the signal amplitude during the eclipse crossing of a radio path. For NAA only, the second '+' or '-'  provides amplitude change for totality on the NAA transmitter, 'nc' means no crossing.}
\end{table}


\section{Results} 

The VLF signals (amplitude and phase) were collected during the day of the eclipse (April 8, 2024), with control measurements of the base level $1-5$ days before and after. The signals collected during the 6 hours of eclipse day are presented in Figure \ref{Fig2}. The signals from the five VLF transmitters were detected with a signal-to-noise ratio of 5 to 50 dB. The receiver in Ann Arbor experienced a high level of noise, allowing only the NAA signal to be reliably observed. 

\begin{figure}[ht!]
\center{\includegraphics[width=.99\textwidth]{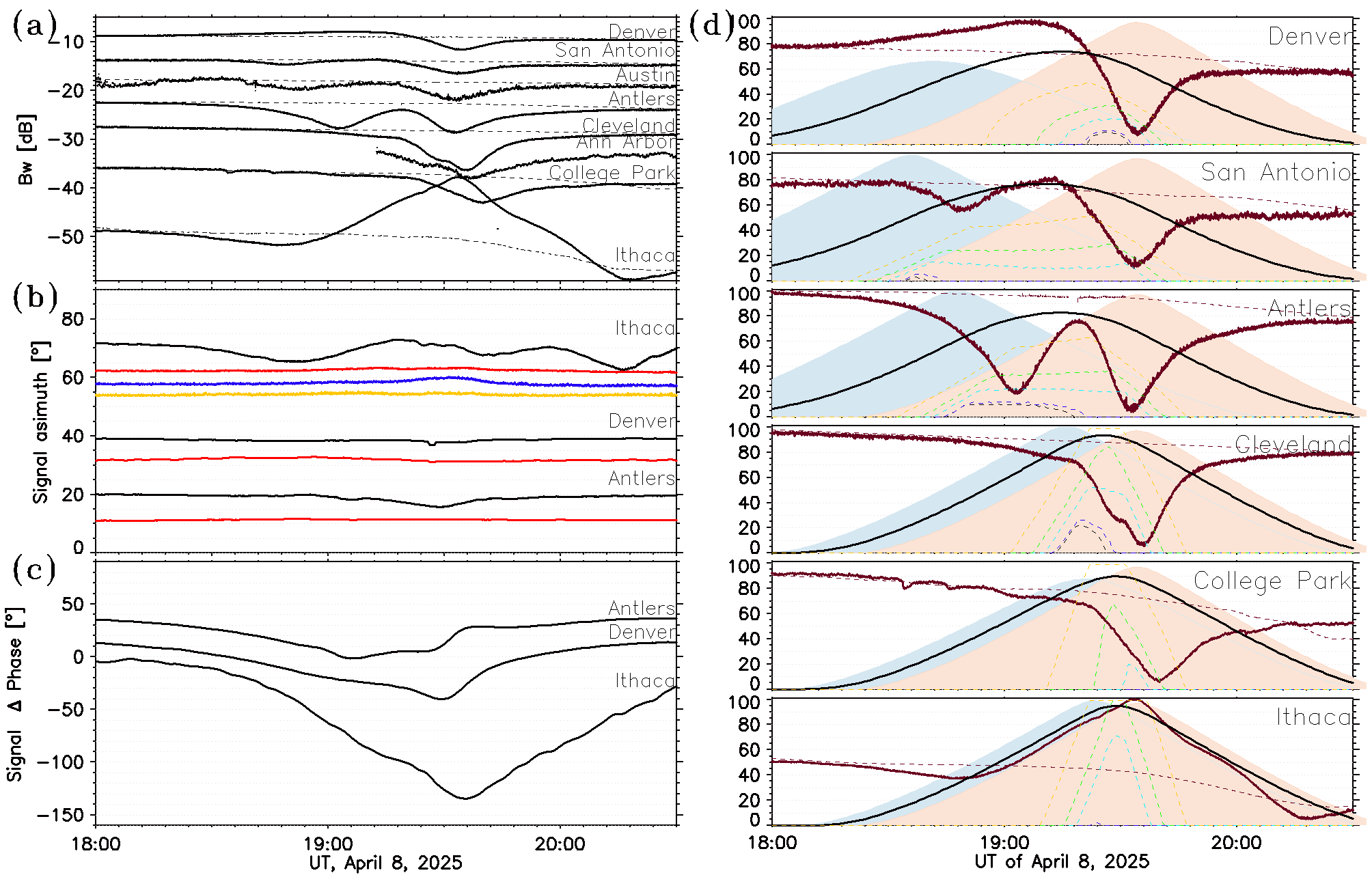}}
 \caption{Dynamics of the signals from the NAA VLF transmitter collected by the VLF receivers during the eclipse passage on April 8, 2024: (a) - the NAA signal observations by the VLF receivers ranged by the distance to the NAA transmitter (from 780 km to Ithaca, NY to 3547 km to Denver, CO) with dashed lines showing baselines from the previous day; (b) - the derived azimuths of NAA (black), NML (red), NLK (blue), and NAU (yellow) signals captured by the VLF receivers in Ithaca, NY, Denver, CO, and Antlers, OK; (c) - the  NAA signal phase perturbation captured by the VFL transmitters in Ithaca, NY, Antlers, OK, and Denver, CO; (d) - the NAA signal observations with the baseline from April 7, 2024 shown by the dashed curves with the transmitter-receiver trace average obscuration (the black solid curves). The obscuration of the NAA transmitter location is highlighted in red; the obscuration of a receiver is highlighted in blue. The fraction of the trace with 100, 99, 95, 90, and 80 \% obscuration is indicated by the dashed curves colored from violet to orange, respectively. 
 }
 \label{Fig3}
\end{figure}

The main focus was on measurements of signal dynamics on radio paths from the closest and most powerful NAA transmitter in Cutler, ME. The NAA transmitter was turned off from 12:00 UT to 17:35 UT, as seen in the amplitude dropout in Figure \ref{Fig2}. For all observations, we focus on the observed signal amplitude, phase, and derived azimuth in the context of simultaneous obscurity at the receiver, the transmitter, and the average obscurity over the entire radio propagation path. An expanded time axis on the eclipse interval for the NAA observations is shown in Figures \ref{Fig3}a and \ref{Fig3}d, where the signals from the receivers are arranged by their distance to NAA, and the timing of the amplitude perturbations relative to the previous day's baseline can be observed. Figure \ref{Fig3}d shows the dynamics of the amplitude overlaid with obscurity curves of the receiver, transmitter, and average radio path. A general decrease in signal amplitude is observed at distances ranging from 998 (College Park) km to 3209 km (San Antonio). This decrease roughly corresponds to the average obscuration of the entire radio propagation path being above 20-30\%, i.e., a duration of approximately 2 hours (indicated in Figure \ref{Fig3}d with the solid black curves).  {A shorter scale (lasting approximately 10-15 minutes) decrease in the signal amplitude, associated with more than 90\% obscuration crossing the path, were observed in San Antonio, Antlers, and Cleveland.}

An increase in signal amplitude is observed at the shortest distance in Ithaca (780 km from NAA): the signal power roughly followed the obscuration of the NAA transmitter. The signal from the farthest receiver in Denver (3547 km) demonstrated a slow, gradual increase of large-scale and a decrease of shorter scale during the transmitter obscuration. 

 {A common feature for all VLF observations (except in Ithaca) is the simultaneous recording of a decrease in the amplitude of the NAA signal during maximum obscurity on the NAA transmitter, with a time scale of approximately 10-15 minutes, which corresponds to a greater than 60-70\% obscuration of the NAA transmitter (Figure \ref{Fig3}d)}.  {This created a characteristic 'W' shape of wave power collected in San Antonio, Austin, and Antlers (the two minima merged in the signal from Cleveland and Ann Arbor).} The receiver in Ithaca recorded a slow 2 dB decrease corresponding to 10-20\% of the radio path obscuration, which turned to a 13 dB 
increase in the amplitude of the NAA signal when the obscuration of the NAA  exceeded 40\%, and was observed simultaneously with the decrease in wave amplitude at distances longer than $\sim1000$ km. 
The derived signal azimuth perturbations of up to 10$^o$ can be seen in Figure \ref{Fig3}b, corresponding to the intensification observed in Ithaca. 

The carrier phase shown in Figure \ref{Fig3}c shows a decrease in the observed NAA carrier phase in all cases. Of special note are the two independent consecutive decreases in amplitude observed in San Antonio and even more prominently at Antlers.  These amplitude valleys are not connected directly to the obscurity curves. Both of these radio propagation paths are nearly aligned with the totality path, and the amplitude decreases are determined by the relative position of the moon's shadow along the path. We return to the significance of these amplitude observations in the discussion section. 

\begin{figure}[ht!]
\center{\includegraphics[width=.99\textwidth]{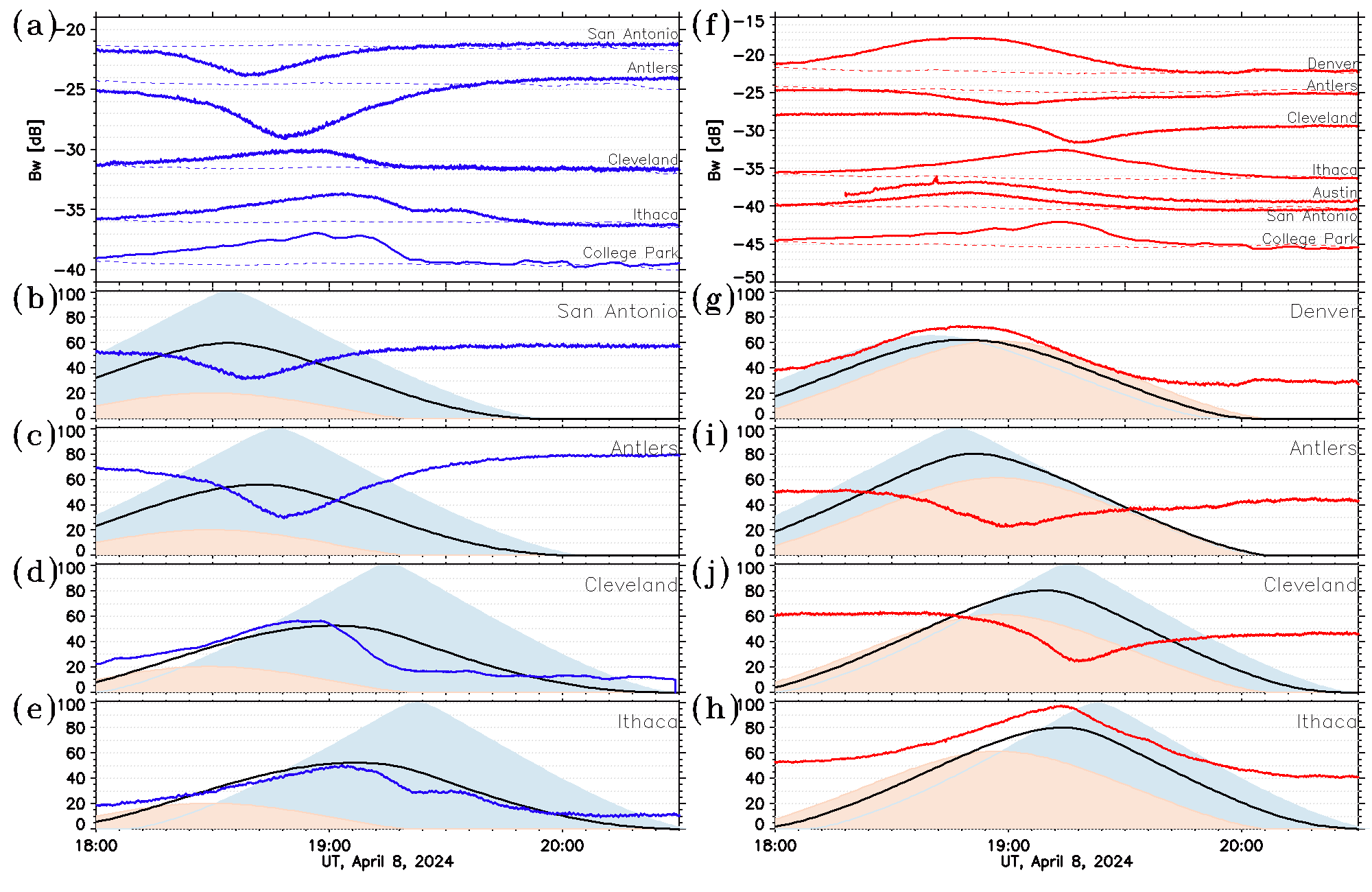}}
 \caption{Dynamics of signal amplitude from the NML (a) and NLK (f) VLF transmitters relative to the previous day (dotted lines )collected by the VLF receivers during the eclipse passage on April 8, 2024, arranged according to the distance to the transmitter. 
(b-e) the obscuration of the NML and (g-h) of the NLK transmitters' location is highlighted in red; the obscuration of a corresponding receiver is highlighted in blue; and the average radio propagation path obscuration is shown with the solid black curve. The recordings by the corresponding receivers (from panels (a) and (f) but with the arbitrary amplitude scale) are shown by solid curves of red (NML) and blue (NLK) colors.  }
 \label{Fig4}
\end{figure}

The amplitude dynamics observed from the NLK transmitter is shown in Figures \ref{Fig4}(a)-(e). Five of the VLF receivers recorded the NLK signal. The receivers, which were closer than 3000 km from the transmitter, in San Antonio (and Austin), and Antlers, OK, detected a $\sim$3 dB decrease of the wave amplitude, even though these locations were on opposite sides of the totality shadow relative to the direction to the transmitter. The receivers in Ithaca, Cleveland, and College Park at distances larger than 3200 km recorded a 2-3 dB signal amplitude enhancement, so the propagation distance is again appears to be a critical parameter. Both receivers in Antlers, OK, and Ithaca, NY, recorded a decrease in signal phase.

The amplitude dynamics observed from the NLM transmitter are shown in Figures \ref{Fig4}(f)-(h). Three regimes of the signal dynamics can be recognized (1) the closest point in Denver (922 km) showed a 5-6 dB increase of the wave amplitude during the maximal obscuration of the propagation path and transmitter of 68\%;  (2) two stations in Antlers and Cleveland ($\sim$1370 and 1440 km) observed a 3-4 dB decrease in the wave amplitude; and (3) stations located more than 1,780 km from the transmitter (specifically in Austin, San Antonio, Ithaca, and Maryland) exhibited a 3-4 dB increase in the signal amplitude. Notably, the receivers in Maryland and Cleveland were positioned along the same NML radio trace.  

\begin{figure}[ht!]
\center{\includegraphics[width=.99\textwidth]{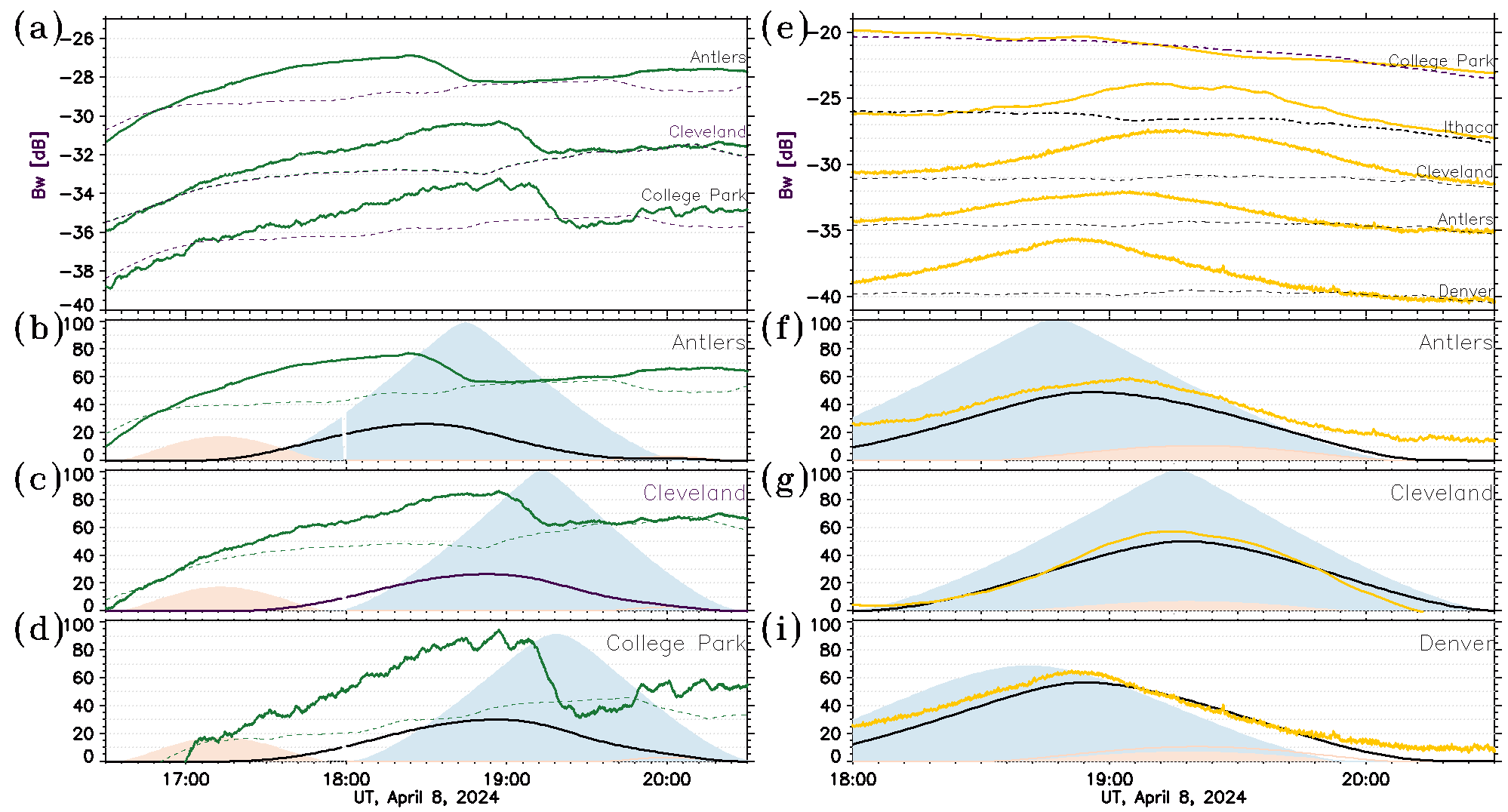}}
 \caption{Dynamics of signal from the NPM (a) and NAU (e) VLF transmitters collected by the VLF receivers during the eclipse passage on April 8, 2024. (b-d) and (f-i) the obscuration of the NPM and  NAU transmitters' location is highlighted in red; the obscuration of a corresponding receiver is highlighted in blue; and the average radio trace obscuration is shown with the solid black curve. The recordings by the corresponding receivers (from panels (a) and (e), but with the arbitrary amplitude scale) are shown by solid curves of green (NPM) and yellow (NAU) colors. 
 }
 \label{Fig5}
\end{figure}

Figure \ref{Fig5} shows the dynamics of the signal amplitude of the NAU and NPM transmitters, which were not locally affected by the solar radiation obscuration during the eclipse. However, the recorded amplitudes from these transmitters did exhibit increases with their maxima timed with the maximum obscuration of the radio paths to the receivers. 

The NPM signal was observed in Cleveland (7337 km, inside the totality), in College Park (7791 km, with the totality crossing the signal path), and in Antlers (6205 km, at the edge of the totality closer to the transmitter).  {All three receivers registered a similar slow increase in signal power by 2-3 dB corresponding to obscuration of the radio path, and a faster decrease of 2-3 dB corresponding to the VLF receivers obscuration increasing above $\sim60-70 \%$ }- the well-known dynamics from the previous eclipse events observations \cite{clilverd_total_2001, cohen_lower_2018, xu_vlf_2019}. A phase decrease of 5-6 degrees was observed. There were no direction changes during the totality crossing.

Five of the VLF receivers (see Table 1) observed the NAU signal (Aguada, Puerto Rico). Four of the receivers exhibited an increase in the observed signal power of 2-4 dB, which is typical for trace lengths between 2000 km and 10000 km \cite{clilverd_total_2001}. The VLF receiver in College Park, which had a maximum obscuration of 89\%, but there was no crossing of the radio trace by the eclipse totality, did not show a significant variation in the NAU signal power. The receivers in Antlers, Ithaca, and Denver recorded a similar phase decrease during the crossing by the totality. { An anticipated reduction in signal power, corresponding to VLF receiver obscuration above 60\% (as observed in the NLK signal power in Figure \ref{Fig5}a-d), was probably masked by the symmetry of the obscuration of the path and transmitter in Cleveland (Figure \ref{Fig5}g). This effect is evident as a time shift of the NAU signal maximum relative to the maximum path obscuration at Antlers (Figure \ref{Fig5}f). }

\section{Discussion}
As noted previously, VLF propagation in the Earth-ionosphere waveguide is multimodal, resulting in nonlinear changes in amplitude and phase during ionospheric perturbations. 
A solar eclipse results in a local decrease in the D-region electron density and a brief transition to conditions that are typically associated with nighttime. This implies an increase in the VLF reflection height and the vertical electron density gradient. The increase in VLF reflection height is associated with a decrease in the VLF carrier phase, which is consistently observed here and in past works.  

The changes in amplitude are more complicated, even though nighttime and eclipsed conditions are associated with lower attenuation. The amplitude at a receiver is the sum of multiple modes, and changes in the modal content from lower attenuation of some modes can lead to total amplitude changes of either polarity. The amplitude of a VLF signal as a function of distance from the source is characterized by a secular decrease along with relatively deep nulls (from destructive interference between modes)   at intervals of several hundred kilometers (see, for example, Figure 6 of \citeA{sasmal2017modeling}). A perturbation of the channel causes a shift of this pattern and the position of the amplitude nulls. So the location of the receiver relative to a null during `ambient' conditions determines the amplitude change at the receiver. An example of this type of sometimes unexpected change can be seen in our observations of the amplitude changes of the NAA signal observed at Ithaca and Cleveland.  The receivers are along the same radio propagation path, with Cleveland 400 km farther away. The NAA signal amplitude at Ithaca increases, while at Cleveland and all other stations it decreases. 
At distances of over 5000 km, VLF propagation tends to be dominated by a single mode, the quasi-TM$_1$ mode, making amplitude changes more predictable. This can be seen in our observations of the NPM and NAU signals, where the amplitudes are seen to increase in all cases, corresponding to lower attenuation during eclipse conditions. 

An alternate explanation of the large $\sim13$ dB increase in amplitude observed at Ithaca and the $\sim7$ dB drop at Cleveland is that there is a significant reflection of the signal and accumulation of wave power within a $\sim$800 km zone during obscuration of the NAA transmitter above 80\%. This can cause an enhancement of the observed NAA signal wave amplitude in the vicinity of the transmitter location and a simultaneous decrease of NAA signal wave power collected outside of this zone. In this context, it is worth noting that reflections from eclipse-induced ionospheric gradients were postulated by \citeA{cohen_lower_2018} 

The most notable amplitude change observation is the NAA amplitude recorded at Antlers (also in Austin and San Antonio, with a lower value of the first minimum), where the amplitude exhibits a `W'-shape as the eclipse passes over the radio propagation path. This means that the amplitude has the same value for 4 different positions of the moon's shadow along the propagation path. 
 {Our estimations based on the timing of the signal from the different stations indicate that the short-scale amplitude signature implies an ionosphere perturbation on a scale of 300 km, which is nearly twice the totality zone. It is worth noting that a 300 km size is roughly on the order of the EUV wavelength 80\% obscuration for eclipse, which is in agreement with the 2017 eclipse estimations in \cite{xu_vlf_2019, golkowski2024vlf}.}
The numerical simulations of these observations have the potential to resolve the size and gradients of the eclipse-induced perturbation.

\section{Conclusions}

We present measurements of VLF signals collected during the total solar eclipse across North America on April 8, 2024. Notably, a powerful radio transmitter was located near the totality shadow, adding a unique aspect to our observations. Our collaborative deployment of nine two-component VLF receivers established a novel multi-point investigation, utilizing a strategic network for comprehensive data collection.

The variation in the propagation of the VLF signal along the paths affected by the eclipse exhibits significant dependence on the path length and the position of the totality zone relative to the receiver, transmitter, and the overall radio path. The central gradual change in amplitude -- either a maximum or minimum -- roughly corresponds to the integrated obscuration along the transmitter-to-receiver path. The first short-scale minimum occurs following approximately 90\% obscuration of the path near the receiver, while the second minimum coincides with the maximum obscuration at the transmitter (NAA) location. The collected VLF data reveal eclipse effects that are distinguishable across different time scales: 

1. The gradual enhancement of signal amplitude up to 5 dB or decrease down to 3 dB, depending on the distance to the transmitter (but farther than 1000 km from the transmitter). This change, along with a slow return to the baseline level, proportionally correlated with the integral obscuration of the entire radio path over a time scale of 2–3 hours. This corresponds to an ionospheric perturbation spatial scale of approximately 2,400 km (roughly the size of the 20\% obscuration zone).; 

 {2. A sharp decrease in signal amplitude lasting 10–15 minutes is associated with the crossing of radio propagation paths of lengths 1000-3000 km by obscuration above 80\% (corresponding to a spatial scale of about 300 km) near the receiver.}

 {3. When the transmitter (NAA) experienced over 80\% obscuration, it caused a sharp, simultaneous decrease in VLF signal amplitude across all detectors for paths longer than 1000 km, occurring on a time scale of 10–15 minutes. This produced a W-shaped signal pattern at distances between 2600 and 3200 km, with overlap minima around 1200–1400 km.}

 {4. The receiver closest to the NAA transmitter, located in Ithaca, NY, recorded the largest change in NAA signal amplitude, an increase of approximately 13 dB during the simultaneous obscuration of the transmitter and receiver above 40\%.}


\section*{Open Research Section}
The data from the experiment used in this paper can be obtained in \cite{agapitov_zenodo2025}. 

\section*{Inclusion in Global Research Statement}
The authors are very grateful for the local help they received from Anna Tenerani of the University of Texas at Austin, Angela Speck of the University of Texas at San Antonio, Natalia Ganushkina of the University of Michigan at Ann Arbor, Kristina Collins of Case Western Reserve University, Jaye Verniero,  Alex Glocer of NASA Goddard Space Flight Center, Czes{\l}aw and Robert Go{\l}kowski in Ithaca, NY.

\acknowledgments
The work is supported by the National Science Foundation with award 2320260 to the University of California Berkeley, and award 2320259 to the University of Colorado Denver both entitled ”Collaborative Research: Remote Sensing of the Lower Ionosphere during 2024 Solar Eclipse: Revealing the Spatial and Temporal Scales of Ionization and Recombination.”

 \bibliography{GRL2025_Eclipse} 

\begin{thebibliography}{}

\bibitem [\protect \citeauthoryear {%
Afraimovich%
, Perevalova%
, Plotnikov%
\BCBL {}\ \BBA {} Uralov%
}{%
Afraimovich%
\ \protect \BOthers {.}}{%
{\protect \APACyear {2002}}%
}]{%
Afraimovich2002}
\APACinsertmetastar {%
Afraimovich2002}%
\begin{APACrefauthors}%
Afraimovich, E\BPBI L.%
, Perevalova, N\BPBI P.%
, Plotnikov, A\BPBI V.%
\BCBL {}\ \BBA {} Uralov, A\BPBI M.%
\end{APACrefauthors}%
\unskip\
\newblock
\APACrefYearMonthDay{2002}{}{}.
\newblock
{\BBOQ}\APACrefatitle {The shock-acoustic waves generated by earthquakes} {The shock-acoustic waves generated by earthquakes}.{\BBCQ}
\newblock
\APACjournalVolNumPages{Annals of Geophysics}{45}{3/4}{471-482}.
\PrintBackRefs{\CurrentBib}

\bibitem [\protect \citeauthoryear {%
Agapitov%
\ \BBA {} Golkowski%
}{%
Agapitov%
\ \BBA {} Golkowski%
}{%
{\protect \APACyear {2025}}%
}]{%
agapitov_zenodo2025}
\APACinsertmetastar {%
agapitov_zenodo2025}%
\begin{APACrefauthors}%
Agapitov, O.%
\BCBT {}\ \BBA {} Golkowski, M.%
\end{APACrefauthors}%
\unskip\
\newblock
\APACrefYearMonthDay{2025}{}{}.
\newblock
{\BBOQ}\APACrefatitle {VLF Transmitters Signal Dynamics from the Radio Paths across and along the Totality Path during the April 8, 2024 Total Solar Eclipse [Data set]} {Vlf transmitters signal dynamics from the radio paths across and along the totality path during the april 8, 2024 total solar eclipse [data set]}.{\BBCQ}
\newblock
\APACjournalVolNumPages{Zenodo}{}{}{}.
\newblock
\begin{APACrefURL} \url{https://doi.org/10.5281/zenodo.15138678} \end{APACrefURL}
\newblock
\begin{APACrefDOI} \doi{10.5281/zenodo.15138678} \end{APACrefDOI}
\PrintBackRefs{\CurrentBib}

\bibitem [\protect \citeauthoryear {%
Albee%
\ \BBA {} Bates%
}{%
Albee%
\ \BBA {} Bates%
}{%
{\protect \APACyear {1965}}%
}]{%
albee_vlf_1965}
\APACinsertmetastar {%
albee_vlf_1965}%
\begin{APACrefauthors}%
Albee, P\BPBI R.%
\BCBT {}\ \BBA {} Bates, H\BPBI F.%
\end{APACrefauthors}%
\unskip\
\newblock
\APACrefYearMonthDay{1965}{{\APACmonth{03}}}{}.
\newblock
{\BBOQ}\APACrefatitle {{VLF} observations at college, {Alaska}, of various \textit{{D}}-region disturbance phenomena} {{VLF} observations at college, {Alaska}, of various \textit{{D}}-region disturbance phenomena}.{\BBCQ}
\newblock
\APACjournalVolNumPages{Planetary and Space Science}{13}{3}{175--206}.
\newblock
\begin{APACrefURL} [{2025-02-14}]\url{https://www.sciencedirect.com/science/article/pii/0032063365900693} \end{APACrefURL}
\newblock
\begin{APACrefDOI} \doi{10.1016/0032-0633(65)90069-3} \end{APACrefDOI}
\PrintBackRefs{\CurrentBib}

\bibitem [\protect \citeauthoryear {%
Baran%
, Shagimuratov%
, Tepenitzina%
\BCBL {}\ \BBA {} Valladares%
}{%
Baran%
\ \protect \BOthers {.}}{%
{\protect \APACyear {2003}}%
}]{%
Baran2003}
\APACinsertmetastar {%
Baran2003}%
\begin{APACrefauthors}%
Baran, L\BPBI W.%
, Shagimuratov, I\BPBI I.%
, Tepenitzina, N\BPBI Y.%
\BCBL {}\ \BBA {} Valladares, C\BPBI E.%
\end{APACrefauthors}%
\unskip\
\newblock
\APACrefYearMonthDay{2003}{}{}.
\newblock
{\BBOQ}\APACrefatitle {Ionospheric response to the solar eclipse of 26 February 1998 over Eastern Europe} {Ionospheric response to the solar eclipse of 26 february 1998 over eastern europe}.{\BBCQ}
\newblock
\APACjournalVolNumPages{Advances in Space Research}{31}{4}{949-954}.
\PrintBackRefs{\CurrentBib}

\bibitem [\protect \citeauthoryear {%
Bonnell%
\ \protect \BOthers {.}}{%
Bonnell%
\ \protect \BOthers {.}}{%
{\protect \APACyear {2021}}%
}]{%
bonnell_first_2021}
\APACinsertmetastar {%
bonnell_first_2021}%
\begin{APACrefauthors}%
Bonnell, J.%
, Agapitov, O.%
, Roglans, R.%
, Samara, M.%
, Robertson, E.%
, McGaw, D.%
\BDBL {}Bortnik, J.%
\end{APACrefauthors}%
\unskip\
\newblock
\APACrefYearMonthDay{2021}{{\APACmonth{12}}}{}.
\newblock
{\BBOQ}\APACrefatitle {First {Results} from {VIPER} {The} {VLF} {Trans}-{Ionospheric} {Propagation} {Experiment} {Rocket} {Campaign}} {First {Results} from {VIPER} {The} {VLF} {Trans}-{Ionospheric} {Propagation} {Experiment} {Rocket} {Campaign}}.{\BBCQ}
\newblock
\APACjournalVolNumPages{}{2021}{}{SA45C--2231}.
\newblock
\begin{APACrefURL} [{2025-03-27}]\url{https://ui.adsabs.harvard.edu/abs/2021AGUFMSA45C2231B} \end{APACrefURL}
\newblock
\APACrefnote{Conference Name: AGU Fall Meeting Abstracts ADS Bibcode: 2021AGUFMSA45C2231B}
\PrintBackRefs{\CurrentBib}

\bibitem [\protect \citeauthoryear {%
Bracewell%
}{%
Bracewell%
}{%
{\protect \APACyear {1952}}%
}]{%
bracewell_theory_1952}
\APACinsertmetastar {%
bracewell_theory_1952}%
\begin{APACrefauthors}%
Bracewell, R\BPBI N.%
\end{APACrefauthors}%
\unskip\
\newblock
\APACrefYearMonthDay{1952}{{\APACmonth{01}}}{}.
\newblock
{\BBOQ}\APACrefatitle {Theory of formation of an ionospheric layer below \textit{{E}} layer based on eclipse and solar flare effects at 16 kc/sec} {Theory of formation of an ionospheric layer below \textit{{E}} layer based on eclipse and solar flare effects at 16 kc/sec}.{\BBCQ}
\newblock
\APACjournalVolNumPages{Journal of Atmospheric and Terrestrial Physics}{2}{4}{226--235}.
\newblock
\begin{APACrefURL} [{2025-02-14}]\url{https://www.sciencedirect.com/science/article/pii/0021916952900330} \end{APACrefURL}
\newblock
\begin{APACrefDOI} \doi{10.1016/0021-9169(52)90033-0} \end{APACrefDOI}
\PrintBackRefs{\CurrentBib}

\bibitem [\protect \citeauthoryear {%
Chakrabarti%
, Sasmal%
, Chakraborty%
, Basak%
\BCBL {}\ \BBA {} Tucker%
}{%
Chakrabarti%
\ \protect \BOthers {.}}{%
{\protect \APACyear {2018}}%
}]{%
chakrabarti2018modeling}
\APACinsertmetastar {%
chakrabarti2018modeling}%
\begin{APACrefauthors}%
Chakrabarti, S\BPBI K.%
, Sasmal, S.%
, Chakraborty, S.%
, Basak, T.%
\BCBL {}\ \BBA {} Tucker, R\BPBI L.%
\end{APACrefauthors}%
\unskip\
\newblock
\APACrefYearMonthDay{2018}{}{}.
\newblock
{\BBOQ}\APACrefatitle {Modeling D-region ionospheric response of the Great American TSE of August 21, 2017 from VLF signal perturbation} {Modeling d-region ionospheric response of the great american tse of august 21, 2017 from vlf signal perturbation}.{\BBCQ}
\newblock
\APACjournalVolNumPages{Advances in Space Research}{62}{3}{651--661}.
\PrintBackRefs{\CurrentBib}

\bibitem [\protect \citeauthoryear {%
Chen%
, Saito%
, Lin%
, Yen%
\BCBL {}\ \BBA {} Tsai%
}{%
Chen%
\ \protect \BOthers {.}}{%
{\protect \APACyear {2021}}%
}]{%
Chen2021}
\APACinsertmetastar {%
Chen2021}%
\begin{APACrefauthors}%
Chen, C\BPBI H.%
, Saito, A.%
, Lin, C\BPBI C\BPBI H.%
, Yen, H\BPBI Y.%
\BCBL {}\ \BBA {} Tsai, H\BPBI F.%
\end{APACrefauthors}%
\unskip\
\newblock
\APACrefYearMonthDay{2021}{}{}.
\newblock
{\BBOQ}\APACrefatitle {Ionospheric response to the solar eclipse of 21 June 2020 over Taiwan} {Ionospheric response to the solar eclipse of 21 june 2020 over taiwan}.{\BBCQ}
\newblock
\APACjournalVolNumPages{Journal of Geophysical Research: Space Physics}{126}{3}{e2020JA028508}.
\PrintBackRefs{\CurrentBib}

\bibitem [\protect \citeauthoryear {%
Clilverd%
\ \protect \BOthers {.}}{%
Clilverd%
\ \protect \BOthers {.}}{%
{\protect \APACyear {2001}}%
}]{%
clilverd_total_2001}
\APACinsertmetastar {%
clilverd_total_2001}%
\begin{APACrefauthors}%
Clilverd, M\BPBI A.%
, Rodger, C\BPBI J.%
, Thomson, N\BPBI R.%
, Lichtenberger, J.%
, Steinbach, P.%
, Cannon, P.%
\BCBL {}\ \BBA {} Angling, M\BPBI J.%
\end{APACrefauthors}%
\unskip\
\newblock
\APACrefYearMonthDay{2001}{}{}.
\newblock
{\BBOQ}\APACrefatitle {Total solar eclipse effects on {VLF} signals: {Observations} and modeling} {Total solar eclipse effects on {VLF} signals: {Observations} and modeling}.{\BBCQ}
\newblock
\APACjournalVolNumPages{Radio Science}{36}{4}{773--788}.
\newblock
\begin{APACrefURL} [{2025-02-13}]\url{https://onlinelibrary.wiley.com/doi/abs/10.1029/2000RS002395} \end{APACrefURL}
\newblock
\APACrefnote{\_eprint: https://onlinelibrary.wiley.com/doi/pdf/10.1029/2000RS002395}
\newblock
\begin{APACrefDOI} \doi{10.1029/2000RS002395} \end{APACrefDOI}
\PrintBackRefs{\CurrentBib}

\bibitem [\protect \citeauthoryear {%
Cohen%
\ \protect \BOthers {.}}{%
Cohen%
\ \protect \BOthers {.}}{%
{\protect \APACyear {2018}}%
}]{%
cohen_lower_2018}
\APACinsertmetastar {%
cohen_lower_2018}%
\begin{APACrefauthors}%
Cohen, M\BPBI B.%
, Gross, N\BPBI C.%
, Higginson-Rollins, M\BPBI A.%
, Marshall, R\BPBI A.%
, Gołkowski, M.%
, Liles, W.%
\BDBL {}Rockway, J.%
\end{APACrefauthors}%
\unskip\
\newblock
\APACrefYearMonthDay{2018}{}{}.
\newblock
{\BBOQ}\APACrefatitle {The {Lower} {Ionospheric} {VLF}/{LF} {Response} to the 2017 {Great} {American} {Solar} {Eclipse} {Observed} {Across} the {Continent}} {The {Lower} {Ionospheric} {VLF}/{LF} {Response} to the 2017 {Great} {American} {Solar} {Eclipse} {Observed} {Across} the {Continent}}.{\BBCQ}
\newblock
\APACjournalVolNumPages{Geophysical Research Letters}{45}{8}{3348--3355}.
\newblock
\begin{APACrefURL} [{2025-02-13}]\url{https://onlinelibrary.wiley.com/doi/abs/10.1002/2018GL077351} \end{APACrefURL}
\newblock
\APACrefnote{\_eprint: https://onlinelibrary.wiley.com/doi/pdf/10.1002/2018GL077351}
\newblock
\begin{APACrefDOI} \doi{10.1002/2018GL077351} \end{APACrefDOI}
\PrintBackRefs{\CurrentBib}

\bibitem [\protect \citeauthoryear {%
Cohen%
, Inan%
\BCBL {}\ \BBA {} Paschal%
}{%
Cohen%
\ \protect \BOthers {.}}{%
{\protect \APACyear {2009}}%
}]{%
cohen2009sensitive}
\APACinsertmetastar {%
cohen2009sensitive}%
\begin{APACrefauthors}%
Cohen, M\BPBI B.%
, Inan, U\BPBI S.%
\BCBL {}\ \BBA {} Paschal, E\BPBI W.%
\end{APACrefauthors}%
\unskip\
\newblock
\APACrefYearMonthDay{2009}{}{}.
\newblock
{\BBOQ}\APACrefatitle {Sensitive broadband ELF/VLF radio reception with the AWESOME instrument} {Sensitive broadband elf/vlf radio reception with the awesome instrument}.{\BBCQ}
\newblock
\APACjournalVolNumPages{IEEE Transactions on Geoscience and Remote Sensing}{48}{1}{3--17}.
\PrintBackRefs{\CurrentBib}

\bibitem [\protect \citeauthoryear {%
Crary%
\ \BBA {} Schneible%
}{%
Crary%
\ \BBA {} Schneible%
}{%
{\protect \APACyear {1965}}%
}]{%
crary1965effect}
\APACinsertmetastar {%
crary1965effect}%
\begin{APACrefauthors}%
Crary, J.%
\BCBT {}\ \BBA {} Schneible, D.%
\end{APACrefauthors}%
\unskip\
\newblock
\APACrefYearMonthDay{1965}{}{}.
\newblock
{\BBOQ}\APACrefatitle {Effect of the eclipse of 20 July 1963 on VLF signals propagating over short paths} {Effect of the eclipse of 20 july 1963 on vlf signals propagating over short paths}.{\BBCQ}
\newblock
\APACjournalVolNumPages{Radio Sci. D}{69}{}{947}.
\PrintBackRefs{\CurrentBib}

\bibitem [\protect \citeauthoryear {%
Dang%
\ \protect \BOthers {.}}{%
Dang%
\ \protect \BOthers {.}}{%
{\protect \APACyear {2020}}%
}]{%
Dang2020}
\APACinsertmetastar {%
Dang2020}%
\begin{APACrefauthors}%
Dang, T.%
, Lei, J.%
, Wang, W.%
, Burns, A.%
, Zhang, B.%
\BCBL {}\ \BBA {} Dou, X.%
\end{APACrefauthors}%
\unskip\
\newblock
\APACrefYearMonthDay{2020}{}{}.
\newblock
{\BBOQ}\APACrefatitle {Ionospheric response to the solar eclipse of 21 June 2020 over China} {Ionospheric response to the solar eclipse of 21 june 2020 over china}.{\BBCQ}
\newblock
\APACjournalVolNumPages{Journal of Geophysical Research: Space Physics}{125}{11}{e2020JA028472}.
\PrintBackRefs{\CurrentBib}

\bibitem [\protect \citeauthoryear {%
Decaux%
\ \BBA {} Gabry%
}{%
Decaux%
\ \BBA {} Gabry%
}{%
{\protect \APACyear {1964}}%
}]{%
decaux1964some}
\APACinsertmetastar {%
decaux1964some}%
\begin{APACrefauthors}%
Decaux, B.%
\BCBT {}\ \BBA {} Gabry, A.%
\end{APACrefauthors}%
\unskip\
\newblock
\APACrefYearMonthDay{1964}{}{}.
\newblock
{\BBOQ}\APACrefatitle {Some particular observations on diurnal phase variations of VLF transmissions received in Paris} {Some particular observations on diurnal phase variations of vlf transmissions received in paris}.{\BBCQ}
\newblock
\APACjournalVolNumPages{Radio Sci}{68}{1}{21--25}.
\PrintBackRefs{\CurrentBib}

\bibitem [\protect \citeauthoryear {%
Gautam%
, MuluyeÂ Tilahun%
, Silwal%
, Adhikari%
\BCBL {}\ \BBA {} GetachewÂ Ejigu%
}{%
Gautam%
\ \protect \BOthers {.}}{%
{\protect \APACyear {2024}}%
}]{%
gautam_ionospheric_2024}
\APACinsertmetastar {%
gautam_ionospheric_2024}%
\begin{APACrefauthors}%
Gautam, S\BPBI P.%
, MuluyeÂ Tilahun, A.%
, Silwal, A.%
, Adhikari, B.%
\BCBL {}\ \BBA {} GetachewÂ Ejigu, Y.%
\end{APACrefauthors}%
\unskip\
\newblock
\APACrefYearMonthDay{2024}{{\APACmonth{10}}}{}.
\newblock
{\BBOQ}\APACrefatitle {Ionospheric response to the 08 {April} 2024 total solar eclipse over {United} {States}: a case study} {Ionospheric response to the 08 {April} 2024 total solar eclipse over {United} {States}: a case study}.{\BBCQ}
\newblock
\APACjournalVolNumPages{Astrophysics and Space Science}{369}{10}{108}.
\newblock
\begin{APACrefURL} [{2025-02-15}]\url{https://doi.org/10.1007/s10509-024-04372-w} \end{APACrefURL}
\newblock
\begin{APACrefDOI} \doi{10.1007/s10509-024-04372-w} \end{APACrefDOI}
\PrintBackRefs{\CurrentBib}

\bibitem [\protect \citeauthoryear {%
Go{\l}kowski%
, Gross%
, Moore%
, Cotts%
\BCBL {}\ \BBA {} Mitchell%
}{%
Go{\l}kowski%
\ \protect \BOthers {.}}{%
{\protect \APACyear {2014}}%
}]{%
golkowski2014observation}
\APACinsertmetastar {%
golkowski2014observation}%
\begin{APACrefauthors}%
Go{\l}kowski, M.%
, Gross, N.%
, Moore, R.%
, Cotts, B.%
\BCBL {}\ \BBA {} Mitchell, M.%
\end{APACrefauthors}%
\unskip\
\newblock
\APACrefYearMonthDay{2014}{}{}.
\newblock
{\BBOQ}\APACrefatitle {Observation of local and conjugate ionospheric perturbations from individual oceanic lightning flashes} {Observation of local and conjugate ionospheric perturbations from individual oceanic lightning flashes}.{\BBCQ}
\newblock
\APACjournalVolNumPages{Geophysical Research Letters}{41}{2}{273--279}.
\PrintBackRefs{\CurrentBib}

\bibitem [\protect \citeauthoryear {%
Golkowski%
\ \protect \BOthers {.}}{%
Golkowski%
\ \protect \BOthers {.}}{%
{\protect \APACyear {2024}}%
}]{%
golkowski2024vlf}
\APACinsertmetastar {%
golkowski2024vlf}%
\begin{APACrefauthors}%
Golkowski, M.%
, Nieckarz, Z.%
, Agapitov, O\BPBI V.%
, Ostrowski, M.%
, Kubisz, J.%
, Mlynarczyk, J.%
\BDBL {}others%
\end{APACrefauthors}%
\unskip\
\newblock
\APACrefYearMonthDay{2024}{}{}.
\newblock
{\BBOQ}\APACrefatitle {VLF and ELF Sensing of the Lower Ionosphere Under Solar Eclipse Conditions} {Vlf and elf sensing of the lower ionosphere under solar eclipse conditions}.{\BBCQ}
\newblock
\APACjournalVolNumPages{AGU24}{}{}{}.
\PrintBackRefs{\CurrentBib}

\bibitem [\protect \citeauthoryear {%
Go{\l}kowski%
, Renick%
\BCBL {}\ \BBA {} Cohen%
}{%
Go{\l}kowski%
\ \protect \BOthers {.}}{%
{\protect \APACyear {2021}}%
}]{%
golkowski2021quantification}
\APACinsertmetastar {%
golkowski2021quantification}%
\begin{APACrefauthors}%
Go{\l}kowski, M.%
, Renick, C.%
\BCBL {}\ \BBA {} Cohen, M.%
\end{APACrefauthors}%
\unskip\
\newblock
\APACrefYearMonthDay{2021}{}{}.
\newblock
{\BBOQ}\APACrefatitle {Quantification of ionospheric perturbations from lightning using overlapping paths of VLF signal propagation} {Quantification of ionospheric perturbations from lightning using overlapping paths of vlf signal propagation}.{\BBCQ}
\newblock
\APACjournalVolNumPages{Journal of Geophysical Research: Space Physics}{126}{5}{e2020JA028540}.
\PrintBackRefs{\CurrentBib}

\bibitem [\protect \citeauthoryear {%
Go{\l}kowski%
\ \protect \BOthers {.}}{%
Go{\l}kowski%
\ \protect \BOthers {.}}{%
{\protect \APACyear {2018}}%
}]{%
golkowski2018ionospheric}
\APACinsertmetastar {%
golkowski2018ionospheric}%
\begin{APACrefauthors}%
Go{\l}kowski, M.%
, Sarker, S.%
, Renick, C.%
, Moore, R.%
, Cohen, M.%
, Ku{\l}ak, A.%
\BDBL {}Kubisz, J.%
\end{APACrefauthors}%
\unskip\
\newblock
\APACrefYearMonthDay{2018}{}{}.
\newblock
{\BBOQ}\APACrefatitle {Ionospheric D region remote sensing using ELF sferic group velocity} {Ionospheric d region remote sensing using elf sferic group velocity}.{\BBCQ}
\newblock
\APACjournalVolNumPages{Geophysical Research Letters}{45}{23}{12--739}.
\PrintBackRefs{\CurrentBib}

\bibitem [\protect \citeauthoryear {%
N.~Gross%
, Cohen%
, Said%
\BCBL {}\ \BBA {} Go{\l}kowski%
}{%
N.~Gross%
\ \protect \BOthers {.}}{%
{\protect \APACyear {2018}}%
}]{%
gross2018polarization}
\APACinsertmetastar {%
gross2018polarization}%
\begin{APACrefauthors}%
Gross, N.%
, Cohen, M.%
, Said, R.%
\BCBL {}\ \BBA {} Go{\l}kowski, M.%
\end{APACrefauthors}%
\unskip\
\newblock
\APACrefYearMonthDay{2018}{}{}.
\newblock
{\BBOQ}\APACrefatitle {Polarization of narrowband VLF transmitter signals as an ionospheric diagnostic} {Polarization of narrowband vlf transmitter signals as an ionospheric diagnostic}.{\BBCQ}
\newblock
\APACjournalVolNumPages{Journal of Geophysical Research: Space Physics}{123}{1}{901--917}.
\PrintBackRefs{\CurrentBib}

\bibitem [\protect \citeauthoryear {%
N\BPBI C.~Gross%
\ \BBA {} Cohen%
}{%
N\BPBI C.~Gross%
\ \BBA {} Cohen%
}{%
{\protect \APACyear {2020}}%
}]{%
gross_vlf_2020}
\APACinsertmetastar {%
gross_vlf_2020}%
\begin{APACrefauthors}%
Gross, N\BPBI C.%
\BCBT {}\ \BBA {} Cohen, M\BPBI B.%
\end{APACrefauthors}%
\unskip\
\newblock
\APACrefYearMonthDay{2020}{}{}.
\newblock
{\BBOQ}\APACrefatitle {{VLF} {Remote} {Sensing} of the {D} {Region} {Ionosphere} {Using} {Neural} {Networks}} {{VLF} {Remote} {Sensing} of the {D} {Region} {Ionosphere} {Using} {Neural} {Networks}}.{\BBCQ}
\newblock
\APACjournalVolNumPages{Journal of Geophysical Research: Space Physics}{125}{1}{e2019JA027135}.
\newblock
\begin{APACrefURL} [{2025-02-13}]\url{https://onlinelibrary.wiley.com/doi/abs/10.1029/2019JA027135} \end{APACrefURL}
\newblock
\APACrefnote{\_eprint: https://onlinelibrary.wiley.com/doi/pdf/10.1029/2019JA027135}
\newblock
\begin{APACrefDOI} \doi{10.1029/2019JA027135} \end{APACrefDOI}
\PrintBackRefs{\CurrentBib}

\bibitem [\protect \citeauthoryear {%
Helliwell%
, Katsufrakis%
\BCBL {}\ \BBA {} Trimpi%
}{%
Helliwell%
\ \protect \BOthers {.}}{%
{\protect \APACyear {1973}}%
}]{%
helliwell1973whistler}
\APACinsertmetastar {%
helliwell1973whistler}%
\begin{APACrefauthors}%
Helliwell, R.%
, Katsufrakis, J.%
\BCBL {}\ \BBA {} Trimpi, M.%
\end{APACrefauthors}%
\unskip\
\newblock
\APACrefYearMonthDay{1973}{}{}.
\newblock
{\BBOQ}\APACrefatitle {Whistler-induced amplitude perturbation in VLF propagation} {Whistler-induced amplitude perturbation in vlf propagation}.{\BBCQ}
\newblock
\APACjournalVolNumPages{Journal of Geophysical Research}{78}{22}{4679--4688}.
\PrintBackRefs{\CurrentBib}

\bibitem [\protect \citeauthoryear {%
Hoque%
, Jakowski%
\BCBL {}\ \BBA {} Berdermann%
}{%
Hoque%
\ \protect \BOthers {.}}{%
{\protect \APACyear {2016}}%
}]{%
Hoque2016}
\APACinsertmetastar {%
Hoque2016}%
\begin{APACrefauthors}%
Hoque, M\BPBI M.%
, Jakowski, N.%
\BCBL {}\ \BBA {} Berdermann, J.%
\end{APACrefauthors}%
\unskip\
\newblock
\APACrefYearMonthDay{2016}{}{}.
\newblock
{\BBOQ}\APACrefatitle {Ionospheric response to the solar eclipse of 20 March 2015 in Europe} {Ionospheric response to the solar eclipse of 20 march 2015 in europe}.{\BBCQ}
\newblock
\APACjournalVolNumPages{Journal of Atmospheric and Solar-Terrestrial Physics}{143-144}{}{45-53}.
\PrintBackRefs{\CurrentBib}

\bibitem [\protect \citeauthoryear {%
Inan%
\ \protect \BOthers {.}}{%
Inan%
\ \protect \BOthers {.}}{%
{\protect \APACyear {2007}}%
}]{%
inan2007massive}
\APACinsertmetastar {%
inan2007massive}%
\begin{APACrefauthors}%
Inan, U\BPBI S.%
, Lehtinen, N\BPBI G.%
, Moore, R.%
, Hurley, K.%
, Boggs, S.%
, Smith, D.%
\BCBL {}\ \BBA {} Fishman, G.%
\end{APACrefauthors}%
\unskip\
\newblock
\APACrefYearMonthDay{2007}{}{}.
\newblock
{\BBOQ}\APACrefatitle {Massive disturbance of the daytime lower ionosphere by the giant $\gamma$-ray flare from magnetar SGR 1806--20} {Massive disturbance of the daytime lower ionosphere by the giant $\gamma$-ray flare from magnetar sgr 1806--20}.{\BBCQ}
\newblock
\APACjournalVolNumPages{Geophysical research letters}{34}{8}{}.
\PrintBackRefs{\CurrentBib}

\bibitem [\protect \citeauthoryear {%
Jakowski%
\ \protect \BOthers {.}}{%
Jakowski%
\ \protect \BOthers {.}}{%
{\protect \APACyear {2008}}%
}]{%
Jakowski2008}
\APACinsertmetastar {%
Jakowski2008}%
\begin{APACrefauthors}%
Jakowski, N.%
, Stankov, S\BPBI M.%
, Wilken, V.%
, Borries, C.%
, Altadill, D.%
, Chum, J.%
\BCBL {}\ \BBA {} Buresova, D.%
\end{APACrefauthors}%
\unskip\
\newblock
\APACrefYearMonthDay{2008}{}{}.
\newblock
{\BBOQ}\APACrefatitle {Ionospheric behavior over Europe during the solar eclipse of 3 October 2005} {Ionospheric behavior over europe during the solar eclipse of 3 october 2005}.{\BBCQ}
\newblock
\APACjournalVolNumPages{Journal of Atmospheric and Solar-Terrestrial Physics}{70}{6}{836-853}.
\PrintBackRefs{\CurrentBib}

\bibitem [\protect \citeauthoryear {%
Kaufmann%
\ \BBA {} Schaal%
}{%
Kaufmann%
\ \BBA {} Schaal%
}{%
{\protect \APACyear {1968}}%
}]{%
kaufmann_effect_1968}
\APACinsertmetastar {%
kaufmann_effect_1968}%
\begin{APACrefauthors}%
Kaufmann, P.%
\BCBT {}\ \BBA {} Schaal, R\BPBI E.%
\end{APACrefauthors}%
\unskip\
\newblock
\APACrefYearMonthDay{1968}{{\APACmonth{03}}}{}.
\newblock
{\BBOQ}\APACrefatitle {The effect of a total solar eclipse on long path {VLF} transmission} {The effect of a total solar eclipse on long path {VLF} transmission}.{\BBCQ}
\newblock
\APACjournalVolNumPages{Journal of Atmospheric and Terrestrial Physics}{30}{3}{469--471}.
\newblock
\begin{APACrefURL} [{2025-02-14}]\url{https://www.sciencedirect.com/science/article/pii/0021916968901190} \end{APACrefURL}
\newblock
\begin{APACrefDOI} \doi{10.1016/0021-9169(68)90119-0} \end{APACrefDOI}
\PrintBackRefs{\CurrentBib}

\bibitem [\protect \citeauthoryear {%
Momani%
, Liu%
\BCBL {}\ \BBA {} Chen%
}{%
Momani%
\ \protect \BOthers {.}}{%
{\protect \APACyear {2010}}%
}]{%
Momani2010}
\APACinsertmetastar {%
Momani2010}%
\begin{APACrefauthors}%
Momani, M\BPBI A.%
, Liu, J\BPBI Y.%
\BCBL {}\ \BBA {} Chen, C\BPBI H.%
\end{APACrefauthors}%
\unskip\
\newblock
\APACrefYearMonthDay{2010}{}{}.
\newblock
{\BBOQ}\APACrefatitle {Ionospheric response to the solar eclipse of 22 July 2009 in Taiwan} {Ionospheric response to the solar eclipse of 22 july 2009 in taiwan}.{\BBCQ}
\newblock
\APACjournalVolNumPages{Journal of Geophysical Research: Space Physics}{115}{A8}{A08304}.
\PrintBackRefs{\CurrentBib}

\bibitem [\protect \citeauthoryear {%
Nieckarz%
\ \protect \BOthers {.}}{%
Nieckarz%
\ \protect \BOthers {.}}{%
{\protect \APACyear {2025}}%
}]{%
nieckarz2024monitoring}
\APACinsertmetastar {%
nieckarz2024monitoring}%
\begin{APACrefauthors}%
Nieckarz, Z.%
, Gołkowski, M.%
, Kubisz, J.%
, Ostrowski, M.%
, Michalec, A.%
, Mlynarczyk, J.%
\BDBL {}Maxworth, A.%
\end{APACrefauthors}%
\unskip\
\newblock
\APACrefYearMonthDay{2025}{}{}.
\newblock
{\BBOQ}\APACrefatitle {Monitoring Global Ionospheric Conditions With Electromagnetic Lightning Impulses Registered in Extremely Low Frequency Measurements} {Monitoring global ionospheric conditions with electromagnetic lightning impulses registered in extremely low frequency measurements}.{\BBCQ}
\newblock
\APACjournalVolNumPages{Radio Science}{60}{2}{e2024RS008140}.
\newblock
\begin{APACrefURL} \url{https://agupubs.onlinelibrary.wiley.com/doi/abs/10.1029/2024RS008140} \end{APACrefURL}
\newblock
\APACrefnote{e2024RS008140 2024RS008140}
\newblock
\begin{APACrefDOI} \doi{https://doi.org/10.1029/2024RS008140} \end{APACrefDOI}
\PrintBackRefs{\CurrentBib}

\bibitem [\protect \citeauthoryear {%
Ostrowski%
\ \protect \BOthers {.}}{%
Ostrowski%
\ \protect \BOthers {.}}{%
{\protect \APACyear {2024}}%
}]{%
ostrowski2024effects}
\APACinsertmetastar {%
ostrowski2024effects}%
\begin{APACrefauthors}%
Ostrowski, M.%
, Go{\l}kowski, M.%
, Kubisz, J.%
, Nieckarz, Z.%
, Michalec, A.%
, Mlynarczyk, J.%
\BDBL {}Maxworth, A.%
\end{APACrefauthors}%
\unskip\
\newblock
\APACrefYearMonthDay{2024}{}{}.
\newblock
{\BBOQ}\APACrefatitle {Effects of a solar flare on global propagation of extremely low frequency waves} {Effects of a solar flare on global propagation of extremely low frequency waves}.{\BBCQ}
\newblock
\APACjournalVolNumPages{Journal of Geophysical Research: Space Physics}{129}{12}{e2024JA033083}.
\PrintBackRefs{\CurrentBib}

\bibitem [\protect \citeauthoryear {%
Resende%
\ \protect \BOthers {.}}{%
Resende%
\ \protect \BOthers {.}}{%
{\protect \APACyear {2021}}%
}]{%
Resende2021}
\APACinsertmetastar {%
Resende2021}%
\begin{APACrefauthors}%
Resende, L\BPBI C\BPBI A.%
, Batista, I\BPBI S.%
, Denardini, C\BPBI M.%
, Carrasco, A\BPBI J.%
, Moro, J.%
, Chen, S\BPBI S.%
\BCBL {}\ \BBA {} Abdu, M\BPBI A.%
\end{APACrefauthors}%
\unskip\
\newblock
\APACrefYearMonthDay{2021}{}{}.
\newblock
{\BBOQ}\APACrefatitle {Ionospheric response to the solar eclipse of 2 July 2019 over South America} {Ionospheric response to the solar eclipse of 2 july 2019 over south america}.{\BBCQ}
\newblock
\APACjournalVolNumPages{Advances in Space Research}{67}{1}{325-335}.
\PrintBackRefs{\CurrentBib}

\bibitem [\protect \citeauthoryear {%
Rishbeth%
}{%
Rishbeth%
}{%
{\protect \APACyear {1968}}%
}]{%
rishbeth1968solar}
\APACinsertmetastar {%
rishbeth1968solar}%
\begin{APACrefauthors}%
Rishbeth, H.%
\end{APACrefauthors}%
\unskip\
\newblock
\APACrefYearMonthDay{1968}{}{}.
\newblock
{\BBOQ}\APACrefatitle {Solar eclipses and ionospheric theory} {Solar eclipses and ionospheric theory}.{\BBCQ}
\newblock
\APACjournalVolNumPages{Space Science Reviews}{8}{4}{543--554}.
\PrintBackRefs{\CurrentBib}

\bibitem [\protect \citeauthoryear {%
Rozhnoi%
\ \protect \BOthers {.}}{%
Rozhnoi%
\ \protect \BOthers {.}}{%
{\protect \APACyear {2020}}%
}]{%
rozhnoi_effect_2020}
\APACinsertmetastar {%
rozhnoi_effect_2020}%
\begin{APACrefauthors}%
Rozhnoi, A.%
, Solovieva, M.%
, Shalimov, S.%
, Ouzounov, D.%
, Gallagher, P.%
, Verth, G.%
\BDBL {}Fedun, V.%
\end{APACrefauthors}%
\unskip\
\newblock
\APACrefYearMonthDay{2020}{}{}.
\newblock
{\BBOQ}\APACrefatitle {The {Effect} of the 21 {August} 2017 {Total} {Solar} {Eclipse} on the {Phase} of {VLF}/{LF} {Signals}} {The {Effect} of the 21 {August} 2017 {Total} {Solar} {Eclipse} on the {Phase} of {VLF}/{LF} {Signals}}.{\BBCQ}
\newblock
\APACjournalVolNumPages{Earth and Space Science}{7}{2}{e2019EA000839}.
\newblock
\begin{APACrefURL} [{2025-02-13}]\url{https://onlinelibrary.wiley.com/doi/abs/10.1029/2019EA000839} \end{APACrefURL}
\newblock
\APACrefnote{\_eprint: https://onlinelibrary.wiley.com/doi/pdf/10.1029/2019EA000839}
\newblock
\begin{APACrefDOI} \doi{10.1029/2019EA000839} \end{APACrefDOI}
\PrintBackRefs{\CurrentBib}

\bibitem [\protect \citeauthoryear {%
Sasmal%
, Basak%
, Chakraborty%
, Palit%
\BCBL {}\ \BBA {} Chakrabarti%
}{%
Sasmal%
\ \protect \BOthers {.}}{%
{\protect \APACyear {2017}}%
}]{%
sasmal2017modeling}
\APACinsertmetastar {%
sasmal2017modeling}%
\begin{APACrefauthors}%
Sasmal, S.%
, Basak, T.%
, Chakraborty, S.%
, Palit, S.%
\BCBL {}\ \BBA {} Chakrabarti, S\BPBI K.%
\end{APACrefauthors}%
\unskip\
\newblock
\APACrefYearMonthDay{2017}{}{}.
\newblock
{\BBOQ}\APACrefatitle {Modeling of temporal variation of very low frequency radio waves over long paths as observed from Indian Antarctic stations} {Modeling of temporal variation of very low frequency radio waves over long paths as observed from indian antarctic stations}.{\BBCQ}
\newblock
\APACjournalVolNumPages{Journal of Geophysical Research: Space Physics}{122}{7}{7698--7712}.
\PrintBackRefs{\CurrentBib}

\bibitem [\protect \citeauthoryear {%
Schaal%
, Mendes%
, Ananthakrishnan%
\BCBL {}\ \BBA {} Kaufmann%
}{%
Schaal%
\ \protect \BOthers {.}}{%
{\protect \APACyear {1970}}%
}]{%
schaal_vlf_1970}
\APACinsertmetastar {%
schaal_vlf_1970}%
\begin{APACrefauthors}%
Schaal, R\BPBI E.%
, Mendes, A\BPBI M.%
, Ananthakrishnan, S.%
\BCBL {}\ \BBA {} Kaufmann, P.%
\end{APACrefauthors}%
\unskip\
\newblock
\APACrefYearMonthDay{1970}{{\APACmonth{06}}}{}.
\newblock
{\BBOQ}\APACrefatitle {{VLF} {Propagation} {Effects} produced by the {Eclipse}} {{VLF} {Propagation} {Effects} produced by the {Eclipse}}.{\BBCQ}
\newblock
\APACjournalVolNumPages{Nature}{226}{5251}{1127--1129}.
\newblock
\begin{APACrefURL} [{2025-02-13}]\url{https://www.nature.com/articles/2261127a0} \end{APACrefURL}
\newblock
\APACrefnote{Publisher: Nature Publishing Group}
\newblock
\begin{APACrefDOI} \doi{10.1038/2261127a0} \end{APACrefDOI}
\PrintBackRefs{\CurrentBib}

\bibitem [\protect \citeauthoryear {%
Tsai%
\ \BBA {} Liu%
}{%
Tsai%
\ \BBA {} Liu%
}{%
{\protect \APACyear {1999}}%
}]{%
Tsai1999}
\APACinsertmetastar {%
Tsai1999}%
\begin{APACrefauthors}%
Tsai, H\BPBI F.%
\BCBT {}\ \BBA {} Liu, J\BPBI Y.%
\end{APACrefauthors}%
\unskip\
\newblock
\APACrefYearMonthDay{1999}{}{}.
\newblock
{\BBOQ}\APACrefatitle {Ionospheric total electron content response to solar eclipses} {Ionospheric total electron content response to solar eclipses}.{\BBCQ}
\newblock
\APACjournalVolNumPages{Journal of Geophysical Research: Space Physics}{104}{A6}{12709-12716}.
\PrintBackRefs{\CurrentBib}

\bibitem [\protect \citeauthoryear {%
Xiong%
\ \BBA {} L{\"u}hr%
}{%
Xiong%
\ \BBA {} L{\"u}hr%
}{%
{\protect \APACyear {2023}}%
}]{%
Xiong2023}
\APACinsertmetastar {%
Xiong2023}%
\begin{APACrefauthors}%
Xiong, C.%
\BCBT {}\ \BBA {} L{\"u}hr, H.%
\end{APACrefauthors}%
\unskip\
\newblock
\APACrefYearMonthDay{2023}{}{}.
\newblock
{\BBOQ}\APACrefatitle {Ionospheric response to solar eclipses as observed by CHAMP satellite} {Ionospheric response to solar eclipses as observed by champ satellite}.{\BBCQ}
\newblock
\APACjournalVolNumPages{Journal of Geophysical Research: Space Physics}{128}{2}{e2022JA031061}.
\PrintBackRefs{\CurrentBib}

\bibitem [\protect \citeauthoryear {%
Xu%
\ \protect \BOthers {.}}{%
Xu%
\ \protect \BOthers {.}}{%
{\protect \APACyear {2019}}%
}]{%
xu_vlf_2019}
\APACinsertmetastar {%
xu_vlf_2019}%
\begin{APACrefauthors}%
Xu, W.%
, Marshall, R\BPBI A.%
, Kero, A.%
, Turunen, E.%
, Drob, D.%
, Sojka, J.%
\BCBL {}\ \BBA {} Rice, D.%
\end{APACrefauthors}%
\unskip\
\newblock
\APACrefYearMonthDay{2019}{{\APACmonth{10}}}{}.
\newblock
{\BBOQ}\APACrefatitle {{VLF} {Measurements} and {Modeling} of the {D}-{Region} {Response} to the 2017 {Total} {Solar} {Eclipse}} {{VLF} {Measurements} and {Modeling} of the {D}-{Region} {Response} to the 2017 {Total} {Solar} {Eclipse}}.{\BBCQ}
\newblock
\APACjournalVolNumPages{IEEE Transactions on Geoscience and Remote Sensing}{57}{10}{7613--7622}.
\newblock
\begin{APACrefURL} [{2025-02-13}]\url{https://ieeexplore.ieee.org/abstract/document/8734830?casa_token=-CV2tiurgo4AAAAA:nXffXdV04Amk6CukzoURh0m1t2oYxLdU0nJjM1gYYfHCg7LdB5fUk9VUvfzF7o7UL1sSRLcqXg} \end{APACrefURL}
\newblock
\APACrefnote{Conference Name: IEEE Transactions on Geoscience and Remote Sensing}
\newblock
\begin{APACrefDOI} \doi{10.1109/TGRS.2019.2914920} \end{APACrefDOI}
\PrintBackRefs{\CurrentBib}

\bibitem [\protect \citeauthoryear {%
Zhang%
\ \protect \BOthers {.}}{%
Zhang%
\ \protect \BOthers {.}}{%
{\protect \APACyear {2017}}%
}]{%
Zhang2017}
\APACinsertmetastar {%
Zhang2017}%
\begin{APACrefauthors}%
Zhang, S\BPBI R.%
, Erickson, P\BPBI J.%
, Goncharenko, L\BPBI P.%
, Coster, A\BPBI J.%
, Rideout, W.%
\BCBL {}\ \BBA {} Vierinen, J.%
\end{APACrefauthors}%
\unskip\
\newblock
\APACrefYearMonthDay{2017}{}{}.
\newblock
{\BBOQ}\APACrefatitle {Ionospheric response to the solar eclipse of 21 August 2017 in the United States} {Ionospheric response to the solar eclipse of 21 august 2017 in the united states}.{\BBCQ}
\newblock
\APACjournalVolNumPages{Geophysical Research Letters}{44}{17}{8664-8671}.
\PrintBackRefs{\CurrentBib}

\end{thebibliography}
\end{document}